# Dynamic Error in Strain Induced Magnetization Reversal of Nanomagnets due to Incoherent Switching and Formation of Metastable States: A Size-dependent Study

Md Mamun Al-Rashid, Student Member, IEEE, Supriyo Bandyopadhyay, Fellow, IEEE, and Jayasimha Atulasimha, Senior Member, IEEE.

*Abstract*—Modulation of stress anisotropy of magnetostrictive nanomagnets with strain offers an extremely energy-efficient method of magnetization reversal. The reversal process, however, is often incoherent and hence error-prone in the presence of thermal noise at room temperature. Occurrence of incoherent metastable states in the potential landscape of the nanomagnet can further exacerbate the error. Stochastic micromagnetic simulations at room temperature are used to understand and calculate energy dissipations and switching error probabilities in this important magnetization switching methodology. We find that these quantities have an intriguing dependence on nanomagnet size: small nanomagnets perform better owing to the fact that they are more resilient to the formation of metastable states and magnetization dynamics in them is more coherent. However, for a fixed stress anisotropy energy density, smaller nanomagnets will also have poorer resilience against thermal instability. Thus, the challenge in straintronics is to maximize the stress anisotropy energy density by developing materials and processes that yield the largest magnetostriction.

## I. Introduction

Nanomagnets acting as binary switches that encode bit information in two stable magnetization orientations [1], [2] are a promising alternative for CMOS based switches. These switches are not only energy-efficient, but also inherently non-volatile, allowing the same device to be used for both logic and memory operations. This opens up avenues for novel computational architectures [3], [4] with improved speed, reliability, functionality and energy saving. The energy-efficiency (how much energy is dissipated in a switching event) of the basic computational unit, i.e. the nanomagnetic switch, however, is critically dependent on the strategy employed to reverse the magnetization. An inefficient strategy will negate the energy advantage over CMOS. A number of techniques to switch (or rotate) the magnetization in nanomagnetic devices have been explored, such as, external magnetic field induced switching [5], spin transfer torque (STT) induced switching implemented by passing an electrical current through a magnetic multilayer [6], [7], stress induced switching of a magnetostrictive nanomagnet brought about by applying an electric potential to an underlying piezoelectric substrate [8]–[11] and spin torque mediated switching due to pure spin current generated by the giant spin Hall effect (SHE) in a heavy metal [12], [13]. Among these strategies, stress induced switching is possibly the most energy efficient. Simulations have shown that a stress clocked dipole coupled nanomagnetic NOT logic gate can be switched in ~1ns with energy dissipation as low as 0.6 aJ and dynamic error probability less than $10^{-8}$ [14]. Estimates based on experimentally demonstrated stress-induced switching of ~200-300 nm lateral dimension elliptical Co and FeGa nanomagnets delineated on a piezoelectric PMN-PT substrate predict that the energy dissipated in the switching process could be as low as a few aJ if the nanomagnets are fabricated on a ~100 nm thin piezoelectric film [11], [15].

This work focuses on studying the stress induced switching strategy which has now been demonstrated experimentally in many systems [11], [15]–[18]. The nanomagnets studied have elliptical shape with in-plane magnetic anisotropy. The nanomagnets have two in plane easy axis (bistable) due to their shape and are expected to exhibit single domain behavior for lateral dimensions of ~100 nm [19]. Elliptical nanomagnets are therefore ideal for encoding binary bits. Magnetic nanostructures with triangular, rectangular and pentagonal shapes have six, four and ten easy axes respectively, and are prone to non-uniformity in the magnetization field and formation of magnetic superstructures, for example vortex [20], which is detrimental to the implementation of nanomagnetic logic. In stress mediated switching of bistable elliptical nanomagnets, magnetization rotation occurs through the modulation of the potential energy landscape of the nanomagnet which allows a maximum rotation of 90º. That becomes problematic if the two stable magnetization states are anti-parallel (angular separation = 180º). After stress withdrawal, the magnetization, which has rotated through $90^0$, may either return to the initial state at $0^0$, or flip to the other stable state (anti-parallel to the first, at $180^0$) with equal probability. Therefore, the probability of magnetization reversal due to stress would be ~50%. There are ways to circumvent this and achieve 180º rotation with probability approaching 100% by suitable engineering [21]–[24] but they introduce additional complexity and are not discussed here. A dipole field, which can be an external magnetic field or a field originating from another nanomagnet, or even a small spin polarized current generating a small spin transfer torque [25], can assist a strained magnetostrictive nanomagnet to flip magnetization and rotate through complete

Submitted on April 11, 2016. M.A. and J.A are supported in part by the National Science Foundation CAREER grant CCF-1253370 and all authors are supported in part by the Commonwealth of Virginia, Center for Innovative Technology (CIT) managed Commonwealth Research Commercialization Fund (CRCF) Matching Funds Program.

M. Al-Rashid and J. Atulasimha are with both the Department of Mechanical and Nuclear Engineering and the Department of Electrical and Computer Engineering, Virginia Commonwealth University, Richmond, VA 23284 USA (e-mail: alrashidmm@vcu.edu, jatulasimha@vcu.edu).

S. Bandyopadhyay is with the Department of Electrical and Computer Engineering, Virginia Commonwealth University, Richmond, VA23284 USA (e-mail: sbandy@vcu.edu)

180º with high probability. In this study, we use a dipole field to achieve $180^0$ rotation in a strained magnetostrictive nanomagnet. The dipole field is just a convenient tool. Strain does the bulk of the work. It rotates the magnetization by $90^0$ and the dipole field then merely tips the balance in favor of $180^0$ rotation (over a $0^0$ rotation), causing complete magnetization reversal with very high probability.

We have performed micromagnetic simulations using MuMax3[26] for a comprehensive understanding of the strain induced switching dynamics i.e. coherency/incoherency of the switching process, its dependence on the nanomagnet dimension, and its influence on the switching reliability and energy dissipation. Stress induced magnetization switching has been shown to be incoherent [27], which motivates our study.

TABLE I: Simulated nanomagnet dimensions.

| Nanomagnet | Length (nm) | Width (nm) | Thickness (nm) |
|---|---|---|---|
| Small | 60 | 40 | 6 |
| Intermediate | 120 | 80 | 12 |
| Large | 150 | 120 | 15 |

## II. MODELING AND SIMULATION

In this study, elliptical disk nanomagnets of three different dimensions have been simulated while keeping the aspect ratio (ratio of major to minor axis to the thickness) constant. A constant aspect ratio ensures constant demagnetizing factors across all dimensions, so the outcome of the simulations will be solely affected by the nanomagnet size. The simulated nanomagnet dimensions are listed in Table I, where "length" is the dimension of the major axis and "width" is that of the minor axis.

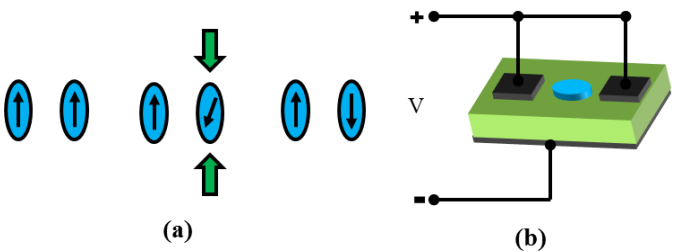

Fig. 1: (a) Stress induced magnetization reversal in the presence of a dipole field caused by a neighboring (left) hard nanomagnet with fixed magnetization. (b) Schematic diagram of the switching set up. A nanomagnet is delineated on top of a piezoelectric substrate and a potential applied between two shorted top electrodes and a bottom electrode generates stress in the nanomagnet inducing its magnetization to rotate [16].

To observe the effect of incoherent reversal in strain induced switching, we use dipole coupling as shown in Fig. 1(a) to ensure complete magnetization reversal since strain, by itself, can produce no more than 90º rotation. This scheme, where a dipole field is utilized to implement the complete reversal under stress, actually represents a very important case, namely the operation of a NOT gate. The neighboring hard nanomagnet's magnetization represents the input bit written into it by some external agent (e.g. a local magnetic field or spin polarized current). The magnetization of the test (soft) nanomagnet represents the output bit. Dipole coupling between the two nanomagnets will prefer to align their magnetizations in mutually anti-parallel orientations, making the output bit the logic complement of the input bit (NOT operation). However, when the input bit is altered, the output bit does not respond and flip automatically to complete the NOT operation since dipole coupling is usually not strong enough to overcome the shape anisotropy of the test nanomagnet and cause magnetization reversal. Therefore, stress is applied to the test nanomagnet by delineating it on top of a piezoelectric layer and applying an electrical voltage on the piezoelectric layer as shown in Fig. 1(b). The generated stress overcomes the shape anisotropy and rotates the test nanomagnet's magnetization by +90º. Later, after stress is withdrawn, the dipole coupling that is always present takes over and preferentially causes another +90º rotation (as opposed to -90º rotation) to flip the magnetization of the test nanomagnet with very high probability. The probability can be made higher by appropriately shaping the stress pulse [28] but those issues are beyond the scope of this paper and not discussed further. This simple reversal process, however, is complicated by random thermal noise at room temperature and incoherency of the magnetization rotation. Stress can also spawn metastable magnetization states in the test nanomagnet and the magnetization vector can get trapped in such a state. Once trapped, it cannot be dislodged by either additional stress or thermal noise. That would cause failure of reversal or switching error. We study all this as a function of nanomagnet size.

TABLE II: Material properties of Terfenol-D [29]–[32].

| Magnetic properties | Terfenol-D |
|---|---|
| Exchange Stiffness ($A_{ex}$) | $9\times10^{-12}$ J/m |
| Saturation Magnetization ($M_s$) | $8\times10^{5}$ A/m |
| Magnetic Exchange Length ($l_{ex}$) | 4.73 nm |
| Gilbert Damping Constant (α) | 0.1 |
| Saturation Magnetostriction ($3/2\,\lambda_s$) | $9\times10^{-4}$ |

Micromagnetic simulation of the switching phenomenon was performed using MuMax3 [26]. Cell sizes less than the magnetic exchange length have been used for discretization in the micromagnetic simulations. When studying stress-induced switching, we assume that the nanomagnets are composed of Terfenol-D which is among the materials with the highest magnetostriction and therefore preferred for straintronic applications since it requires lower stress for switching. The material parameters are listed in Table II. This material is chosen to minimize the energy dissipation in stress-induced switching and has been grown successfully by others [32]. The explicit form for the Landau-Lifshitz torque used by MuMax3 is [26]

$$\frac{d\vec{m}}{dt} = \vec{\tau}_{LL} = \gamma_{LL} \frac{1}{1+\alpha^2}(\vec{m} \times \vec{H}_{eff} + \alpha(\vec{m} \times (\vec{m} \times \vec{H}_{eff}))) \quad (1)$$

Here, $\gamma_{LL}$ is the gyromagnetic ratio (rad/Ts), $\alpha$ is the dimensionless Gilbert damping parameter and $\vec{H}_{eff}$ is effective field,

$$\vec{H}_{eff} = \vec{H}_{ext} + \vec{H}_{demag} + \vec{H}_{exch} + \vec{H}_{anis} + \vec{H}_{therm} \quad (2)$$

where, $\vec{H}_{ext}$ is the externally applied field, $\vec{H}_{demag}$ is the magnetostatic field, $\vec{H}_{exch}$ is the exchange field, $\vec{H}_{anis}$ is the magneto-crystalline anisotropy field (which includes uniaxial and cubic anisotropy) and, $\vec{H}_{therm}$ is the random thermal field representing thermal noise.

In dipole coupled nanomagnetic NOT logic as shown in Figure 1(a), the effective dipole field experienced by one magnet from another can be incorporated as an external field $\vec{H}_{ext}$ [33]. To incorporate the effect of stress, the uniaxial anisotropy field has been exploited. Uniaxial magneto-crystalline anisotropy is modeled in MuMax3 using the following effective field term

$$\vec{H}_{anis} = \frac{2K_{u1}}{\mu_0 M_{sat}}(\vec{u}.\vec{m})\vec{u} + \frac{4K_{u2}}{\mu_0 M_{sat}}(\vec{u}.\vec{m})^3 \vec{u} \quad (3)$$

where, $K_{u1}$ and $K_{u2}$ are first and second order uniaxial anisotropy constants, $M_{sat}$ is the saturation magnetization and $\vec{u}$ is the unit vector in the direction of the anisotropy. Assigning $K_{u2} = 0$, Equation (3) reduces to

$$\vec{H}_{anis} = \frac{2K_{u1}}{\mu_0 M_{sat}}(\vec{u}.\vec{m})\vec{u} \quad (4)$$

The effective field due to an external uniaxial stress can be expressed as [34]

$$\vec{H}_{stress} = \frac{3\lambda_s \sigma}{\mu_0 M_{sat}}(\vec{s}.\vec{m})\vec{s} \quad (5)$$

where, $(3/2)\lambda_s$ is the saturation magnetostriction, $\sigma$ is the external stress (Pa) and $\vec{s}$ is the unit vector in the direction of the applied stress. Comparing Equations (4) and (5), the following equation can be used to find the value of $K_{u1}$ to effectively simulate the effect for a given uniaxial stress $\sigma$ applied in the same direction as the uniaxial anisotropy as

$$K_{u1} = \frac{3\lambda_s \sigma}{2} \quad (6)$$

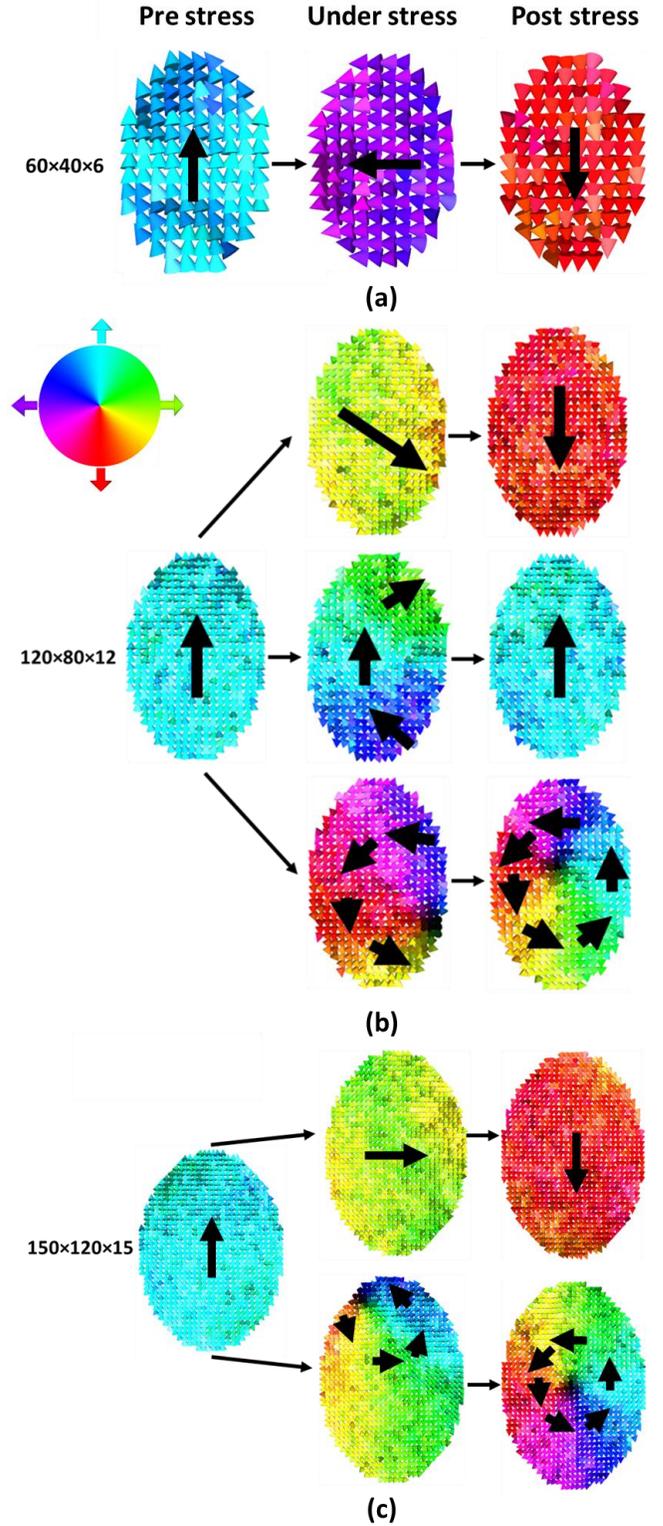

Fig. 2: Strain mediated reversal. Nanomagnet dimensions, (a) Small, (b) Intermediate, (c) Large.

## III. RESULTS

To observe the coherency/incoherency of stress mediated switching, micromagnetic simulations have been performed at room temperature (300 K) by including a random thermal field in the $H_{eff}$ term. The switching coherency/incoherency is dependent on the *size of the nanomagnet* as can be seen in Fig. 2.

The stress applied on the three nanomagnets of dimensions given in Table I is equal to the critical stress. Critical stress is defined as the stress for which the stress anisotropy energy equals the shape anisotropy energy. It is independent of the magnet's volume, but depends on the magnet's aspect ratio. Since all three nanomagnets has the same aspect ratio, they have equal shape anisotropy energy density and therefore, equal critical stress. The incoherency in the switching process clearly increases with increasing dimension. Exchange interaction forces the spins to rotate together in unison in smaller nanomagnets resulting in a coherent rotation in the smallest nanomagnet as shown in Fig. 2(a). The magnetization rotates completely to the opposite direction after stress withdrawal. Increasing the dimensions allows spins to reduce the exchange energy penalty be amortizing over a larger number of spins, resulting in incoherent rotation in nanomagnets with larger dimensions. Because of the incoherency in the switching process, two incoherent metastable states, namely the C-state and vortex state are found in the larger nanomagnets when stress is applied. The C-state is so named since in this state the spins seem to arrange themselves in the form of the letter C (the spin texture curls to form the shape of the letter C). In the intermediate sized nanomagnet, the possible outcomes after the application of stress as shown in Fig. 2(b) are: (1) the magnetization can go through a nearly coherent rotation and emerge in the opposite direction, thus completing the switching process, (2) the magnetization can go to the C-state until the stress is withdrawn and come back to the initial state after stress withdrawal or (3) the magnetization can enter the vortex state and remain there even after stress withdrawal. Among these two incoherent metastable states, the C-state is the most prominent in the intermediate sized nanomagnet while rarely entering the vortex state (< 1% probability). On the other hand, the C-state is completely absent in the large nanomagnet. This magnet either switches successfully or enters the vortex state and remains there as shown in Fig. 2(c).

Switching error calculation has been performed at room temperature (300 K). The thermal field causes erratic magnetization rotation and hence the magnetization may fail to rotate through 180º, resulting in a switching error. The switching error has been estimated from simulations of 1000 switching trajectories (for each case). Stress is turned on at some time and we follow the temporal evolution of the magnetization perturbed by a random thermal field (mimicking thermal noise) at every simulation time step. This generates a switching trajectory in MuMax3 simulation. The fraction of the trajectories that fail to switch by completing a ~180º rotation is the error probability. *We deliberately chose parameters where the switching error is 1% or higher*, so that only 1000 micromagnetic simulations would suffice for generating the required statistics. To study switching error probabilities < 1% we would need to increase the number of trajectories but we do not study that regime as it is computationally prohibitive and more importantly the physics and trends observed at errors rates over 1% would also scale to that regime. Hence, studying the low error probability regime would have not only stretched our resources but also would have been superfluous within the scope of this paper.

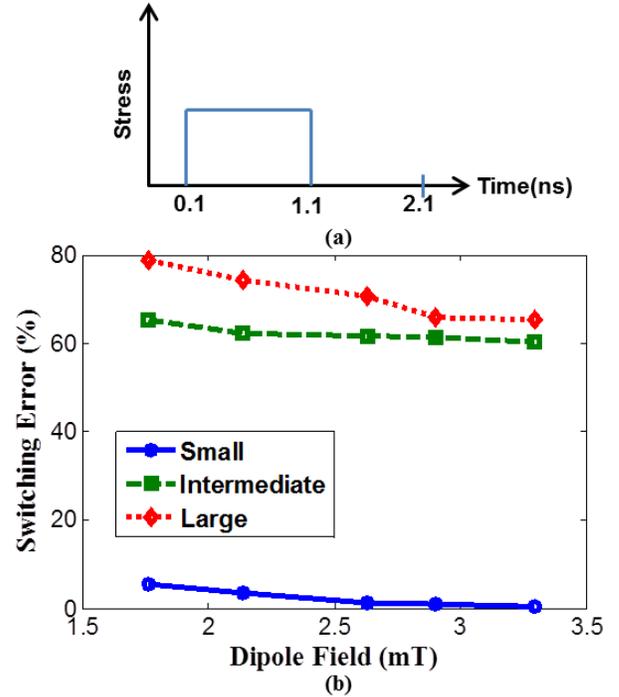

Fig. 3: Strain mediated switching. a) Stress profile in time domain. b) Dipole field strength vs. percentage of switching error in stress mediated switching.

Fig. 3 (a) shows the stress profile used in our simulations and was applied to all the nanomagnets. The voltage required to generate the applied stress (~28 MPa) for a 100 nm thick PZT thin film using the scheme shown in Fig. 1(b) is ~105 mV and the ferroelectric response can be considered instantaneous. (The ferroelectric response has been demonstrated to have a characteristic switching time constant of ~70-90 ps [35]). Fig. 3(b) shows the estimated switching error probability as a function of varying dipole coupling field. The results show that the switching error probability decreases with increasing dipole field for all nanomagnet dimensions. This is expected since the dipole field favors the +90º rotation (right) over the -90º rotation (wrong) after stress removal and hence increasing its strength would reduce error probability. However, the switching error probability also increases with increasing nanomagnet size and this is due to increasing

5incoherency in the switching process which makes magnetization dynamics increasingly vulnerable to thermal noise in larger nanomagnets.

There is another effect that is responsible for making larger nanomagnets more error-prone. In larger nanomagnets, stress spawns metastable states (C-state and vortex state) in the nanomagnet's potential energy profile. They trap the magnetization and impede successful switching. Neither additional stress, nor thermal noise, can easily dislodge the magnetization from the metastable state. As shown in Fig. 2, the metastable states can have a spin texture that is either C shaped (present only in the intermediate dimension) or a vortex state (present in both intermediate and large nanomagnets), both of which are highly incoherent states and therefore never present in the small nanomagnet. This explains the high switching error in the intermediate and large nanomagnets compared to the small one.

It should be noted that the C-state, once formed, is sustained in the presence of stress but it still has net magnetization predominantly pointing in the initial direction. Hence, when stress is withdrawn, it always returns the magnetization to the initial state, i.e. the nanomagnet does not switch. The vortex state has a net magnetization equal to zero and has no memory of the initial state. However, it too prevents switching because it is so stable that the magnetization remains in this state even after stress is withdrawn. The difference is that that unlike the C-state, the vortex state does not return the magnetization to the initial orientation upon stress withdrawal, but traps it into a different orientation. The energy barrier that surrounds the metastable vortex state cannot be overcome by the dipole coupling or thermal fluctuations at room temperature. Nor can stress (even ~200 MPa) destroy this state, once formed. Thus, the only way intermediate and large nanomagnets can reverse their magnetization successfully is by altogether avoiding the C- and vortex-states. That is why the switching error probabilities are very high for intermediate and large nanomagnets in Fig. 3(b).

In the larger nanomagnet, the probability of formation of the vortex state decreases with increasing dipole coupling that results in lower switching error as the dipole coupling is increased. This reduced probability of formation of the vortex state is also true in intermediate nanomagnets but the reduction in the probability of formation of the C-state is comparatively smaller in the range of dipole fields studied. Thus, the decrease in switching error with increasing dipole coupling is less pronounced in the intermediate sized nanomagnet compared to the large one as seen in Fig. 3(b).

To verify that the observed trend in the switching error rates with respect to dipole field as shown in Fig. 3(b) continues to the regime with error rates of <1%, we have simulated 10000 trajectories for the small (60nm - 40nm - 6nm) nanomagnet with dipole fields of 5.44mT and 7.23 mT under the same stress profile as shown in Fig. 3(a). The simulations show 5 and 2 errors respectively out of the 10000 trajectories (0.05% and 0.02% error probability) whereas there were 5 errors out of 1000 trajectories for a dipole field of 3.3 mT (0.5%). So, the error rates are indeed scaling.

## IV. Conclusion

This study shows that strain based switching is coherent in small nanomagnets (~60 nm × 40 nm × 6 nm) and can be switched reliably. The switching gets more incoherent as the nanomagnets size increases, leading to large errors in switching. All this indicates that straintronic switching is best suited for small nanomagnets (lateral dimensions ~ 50 nm or smaller) that could result in reliable switching and extremely energy efficient operation. However, there is one caveat. Normally, the shape anisotropy barrier has to be sufficiently tall (~ 1.7 eV) to ensure enough thermal stability. The stress anisotropy energy has to equal or exceed the barrier height in order to cause the $90^0$ rotation that is necessary for the reversal. The stress anisotropy energy is the product of the stress, the magnetostriction coefficient, and the nanomagnet volume. In smaller nanomagnets, we will require either a larger stress (undesirable since it increase the switching voltage and energy dissipation and in some cases impossible to apply as the strain that the piezoelectric can generated in limited) or a larger magnetostriction coefficient. Terfenol-D with saturation magnetostriction of ~1500 ppm [36] can be used in fabricating small nanomagnets. While this material has indeed been sputtered to produce thin films with saturation magnetostriction of ~900 ppm [31], fabricating small nanomagnets with this material proves to be challenging. On the other hand, materials like Co or Ni that have been used to demonstrate strain switched magnetostrictive nanomagnets [11], [16] have saturation magnetostriction of only ~ 30 ppm. Using Ni/Co nanomagnets, the stress anisotropy energy achievable at dimensions ~60 nm×40 nm×6 nm with a stress as large as 100 MPa is a mere ~0.2 eV (~ 8kT at room temperature) which is way too low to ensure sufficient thermal stability. There is a limit on the amount of stress that can be generated or sustained. That precludes nanomagnets that are simultaneously small, amenable to switching by stress and possess high stability against thermal noise at room temperature. Larger nanomagnets would solve this issue as increasing the nanomagnet volume while retaining the same stress anisotropy energy density allows the generation of larger stress anisotropy energy and hence a larger shape anisotropy barrier for better thermal stability. Unfortunately, as this study shows, larger nanomagnets are considerably more error-prone. This clearly demonstrates the challenges that one faces in straintronics, namely the design of nanomagnets with sufficient thermal stability and sufficient resilience against switching errors. Clearly, a large magnetostriction always helps and hence there is an urgent need to develop processes and materials that yield high magnetostriction.

## Appendix

### I. Energy Dissipation estimates in stress mediated switching

To estimate the total energy dissipation in the stress mediated switching process, two energy dissipation mechanisms have been considered, 1) The internal energy dissipation within the nanomagnet due to Gilbert damping and 2) circuit energy dissipation in charging and discharging the capacitive PZT layer. These two energies are added together to get the total energy dissipation. The PZT film is considered to be 100nm thick. Total energy dissipation in the case of the smallest nanomagnet (60nm×40nm×6nm) for achieving 1% switching error has been estimated to be 5.5 aJ. The estimated energy dissipation for stress mediated switching at different dipole fields is listed in Table A-I.

Table A-I. Energy dissipation in stress mediated switching.

| Nanomagnet Size | Energy dissipation ($\times 10^{-18}$ J) for dipole fields (mT) | | | | |
|---|---|---|---|---|---|
| | *1.7630* | *2.1400* | *2.6320* | *2.9* | *3.294* |
| Small | 5.4860 | 5.4894 | 5.4938 | 5.4962 | 5.4998 |
| Intermediate | 5.5976 | 5.6249 | 5.6605 | 5.6799 | 5.7084 |
| Large | 5.7691 | 5.8330 | 5.9165 | 5.9619 | 6.0288 |

The switching reliability can be increased significantly (switching error << 1%) in the case of the smallest nanomagnet by increasing the dipole field and/or increasing the stress application time. Increasing the dipole field causes very small increase in the energy dissipation in the form of damping as can be seen from Table A-I. Increasing the time over which stress is applied makes the device slower but does not change the energy dissipation. So, the switching reliability can be significantly improved with very small penalty in terms of energy dissipation.

## Acknowledgment

M.A. and J.A are supported in part by the National Science Foundation CAREER grant CCF-1253370 and all authors are supported in part by the Commonwealth of Virginia, Center for Innovative Technology (CIT) managed Commonwealth Research Commercialization Fund (CRCF) Matching Funds Program.

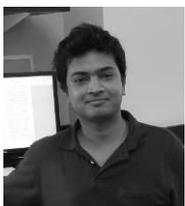

**Md Mamun Al-Rashid** received the B.Sc. degree in electrical and electronic engineering from the Bangladesh University of Engineering and Technology, Dhaka, Bangladesh, in 2012. He is currently pursuing the Ph.D. degree in electrical and computer engineering with Virginia Commonwealth University, Richmond, VA, USA. He is also affiliated with the Department of Mechanical and Nuclear Engineering, Virginia Commonwealth University.

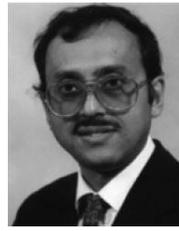

**Supriyo Bandyopadhyay** (SM'80–M'86–SM'88– F'05) is currently a Professor of Electrical and Computer Engineering with Virginia Commonwealth University, Richmond, VA, USA, where he is the Director of the Quantum Device Laboratory. He has authored or co-authored over 300 scientific publications. His current research interests include nanoelectronics, nanomagnetism, and spintronics.

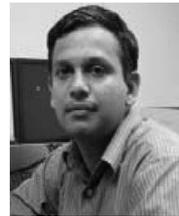

**Jayasimha Atulasimha** (SM'11) received the M.S. and Ph.D. degrees in Aerospace Engineering from the University of Maryland, College Park, MD, USA, in 2003 and 2006, respectively. He is an Associate Professor of Mechanical and Nuclear Engineering and also has a courtesy appointment as Associate Professor of Electrical and Computer Engineering at the Virginia Commonwealth University, Richmond, VA, USA where he directs the Magnetism, Magnetic Materials and Magnetic Devices (M3) laboratory. He has authored or coauthored about 50 journal publications on magnetostrictive materials, magnetization dynamics, and nanomagnetic computing. His research interests include magnetostrictive materials, nanoscale magnetization dynamics, and multiferroic nanomagnet-based computing architectures.

Prof. Atulasimha received the NSF CAREER Award for 2013–2018.